\documentclass[10pt]{iopart}

\usepackage{graphicx}

\begin{document}

\title[Emulsions with variable interfacial tension]
{Colloid-stabilized
emulsions: behaviour as the interfacial tension is reduced}

\author{P.S.~Clegg\dag \S, E.M.~Herzig\dag \S, A.B.~Schofield\dag, 
T.S.~Horozov\ddag,
B.P.~Binks\ddag, M.E.~Cates\dag, and W.C.K.~Poon\dag \S}

\address{\dag\ SUPA, School of Physics, University of
Edinburgh, Edinburgh, EH9 3JZ, United Kingdom}

\address{\S\ COSMIC, University of Edinburgh, Edinburgh, EH9 3JZ, United
Kingdom}

\address{\ddag\ Surfactant \& Colloid Group, Department of Chemistry,
University of Hull, Hull, HU6 7RX, United Kingdom}

\begin{abstract}
We present confocal microscopy studies of novel particle-stabilized
emulsions. The novelty arises because the immiscible fluids have an 
accessible upper
critical solution temperature. The emulsions have been created by
beginning with particles dispersed in the single-fluid phase. On
cooling,
regions of the minority phase nucleate. While coarsening these nuclei
become coated with particles due to the associated reduction in
interfacial energy. The
resulting emulsion is arrested, and the particle-coated interfaces have
intriguing properties. Having made use of the binary-fluid phase diagram
to create the emulsion we then make use of it to study the properties of
the interfaces. As the emulsion is re-heated toward the single-fluid 
phase the
interfacial tension falls and the volume of the dispersed phase drops.
Crumpling, fracture or coalescence can follow. The results show that the
elasticity of the interfaces has a controlling influence over the
emulsion behaviour.
\end{abstract}


\section{Introduction}

Recent studies have begun to examine the mechanism by which colloidal
particles can emulsify immiscible liquids. One strand of 
this research has focused on
different oils emulsified in water and has used silica
particles as the prototype colloid~\cite{Aveyard2003}. 
If the colloid is partly wetted by both
fluids
then there is a large energy gain if the colloid resides on the
interface between the two liquids. The benefit arises because the
particle removes a region of the shared interface between the two
fluids. The simplicity of the mechanism and the stability of the
resulting emulsions makes this a very promising
technology.

\begin{figure}
\centerline{\includegraphics[scale=1.2]{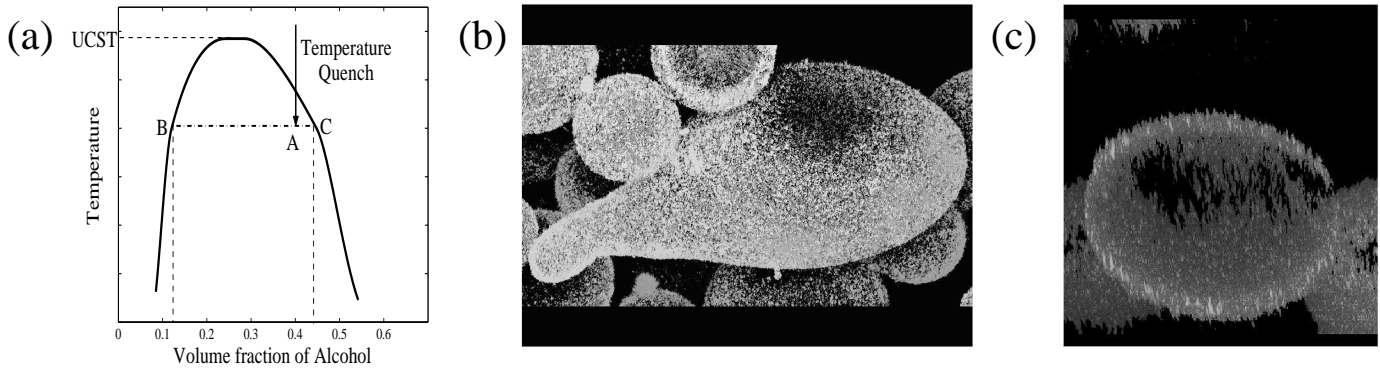}}
\caption{\label{Projection}
(a) A schematic alcohol-oil phase diagram showing the quench route used
here. The bold line is the coexistence curve and the UCST is the upper
critical solution temperature. A sample quenched to A will nucleate
droplets with composition B coexisting with a continuous background of
composition C.
(b \& c) A hexane in methanol emulsion is stabilized by silica particles
following a slow shallow quench into the demixed region. These
images are a projection of colloidal stacks and they show that there is
no systematic distortion due to gravity (pulling them up). The vertical
dimension is 200 $\mu$m.}
\end{figure}

Numerous pairs of immiscible liquids will begin to mix when heated or
cooled~\cite{Rowlinson1982}. The case of liquids with an upper critical
solution temperature (UCST), which mix on warming, is shown in
figure~\ref{Projection}(a).
As the mixed state is approached the compositions of the two
liquids changes steadily
and the interfacial tension between them
decreases.
This effect gives novel opportunities for studying
particle-stabilized emulsions. It makes it possible for us to observe the
behaviour of the droplets as the balance of volumes of the two fluids
changes continuously. In addition the effect of vanishing interfacial
tension on the colloid layer can be observed directly.

In this paper we study alcohol-oil emulsions stabilized by large
silica colloids. The particles themselves are $\sim$0.5 $\mu$m diameter
St\"{o}ber silica~\cite{Stober1968} 
with modified surface chemistry. The surface has been
partially silanized to give a wetting angle close to 90$^{\circ}$ for
the alcohol-oil-silica interface~\cite{Horozov2003}. 
In the systems studied here we have
used methanol-hexane, methanol-heptane and ethanol-dodecane as the
alcohol-oil combinations. For these pairs the upper critical points are
in
a convenient temperature range (12$^{\circ}$ -- 53$^{\circ}$~C). 
The liquid compositions (generally 2:3
alcohol:oil by volume) were away from the
critical composition (see the quench route in figure~\ref{Projection}(a))
and particle volume fractions of $\sim$1\% were used.
The force keeping a large colloidal particle on a liquid-liquid
interface is surprisingly large. Close to the neutral wetting condition, a
$\sim$0.5 $\mu$m
particle can change the interfacial energy by $\sim$10$^4~k_BT$. We used a
Nikon confocal microscope with a Biorad Ar-ion laser to examine the
samples. Rendering 3-dimensional images was carried out using the ImageJ
software package~\cite{ImageJ}.

Our studies were performed by emulsifying
the liquids using the silica particles and observing the changes as the
liquids were warmed toward miscibility. We found
evidence that the particle-stabilized interfaces are solid.
The solid behaviour of particles trapped at flat
liquid-liquid interfaces has previously been studied in
isolation~\cite{Aveyard2000}~\cite{Stancik2004}.
Furthermore, in emulsion
science and technology the case of protein stabilized interfaces has
similarities with the emulsion considered here~\cite{Dickinson1999}. 
Globular proteins often
begin to bond with each other on liquid-liquid interfaces and this
strengthens the resulting emulsion. Here we show the behaviour of
densely-laden interfaces as the interfacial tension falls.

\section{Behaviour of droplets as the interfacial tension changes}

\subsection{Creation by nucleation}

The emulsification route used here is novel.
Many of our emulsions have been created
by beginning with the miscible liquid pair and the colloids dispersed in
this single fluid phase. The interfaces and concomitant interfacial
tension only appear as the sample is quenched into the two-phase region,
figure~\ref{Projection}(a).
Emulsions form on
cooling as particles are swept up by the interfaces of the expanding
droplets. The layer
of colloids at the interface eventually becomes very dense. The droplets
stop growing and are stable for many hours, figure~\ref{Projection}(b \& c). 
The final
length scale of the structure is controlled by the geometry of surface
coverage. The droplet diameter is $\xi \sim n R / \phi_v$
where $R$ is the colloid radius, $\phi_v$ is their volume fraction, and
$n$ is the number of layers of particles.
In the samples studied here the interfaces are covered by half-a-dozen
layers of particles and this is consistent with the observation that 
$\xi \sim 150~\mu$m.

A comparison between the
particle-free case and the emulsified droplets created here is
instructive. For alcohol-oil mixtures 
undergoing demixing, gravity will eventually 
cause the droplets
to cream to the top of the sample and coalesce. In
pure alcohol-oil mixtures this leads to macroscopic phase separation.
The response to creaming is evidently strongly
modified by the colloids. The droplets still rise due to gravity;
however they maintain their integrity due to the particle-stabilized
interfaces.
The coalescence mechanism for droplets emulsified by a conventional 
surfactant has
been studied for many years~\cite{Evans1999}. 
The role of fluctuations in the interfacial
coverage and of the equilibrium between surfactant in solution and on
the interface have been shown to be important. These effects are all
associated with the surfactant layer being fluid.
Here the colloidal particles are very densely packed and it
seems unlikely that regions of exposed interface will appear by
fluctuation alone. The dense interface will strongly oppose the
formation of necks between two droplets. Hence, as we observe, the 
colloid-stabilized emulsion is robust to coalescence.

\subsection{Destruction on warming}

\begin{figure}
\centerline{\includegraphics[scale=0.5]{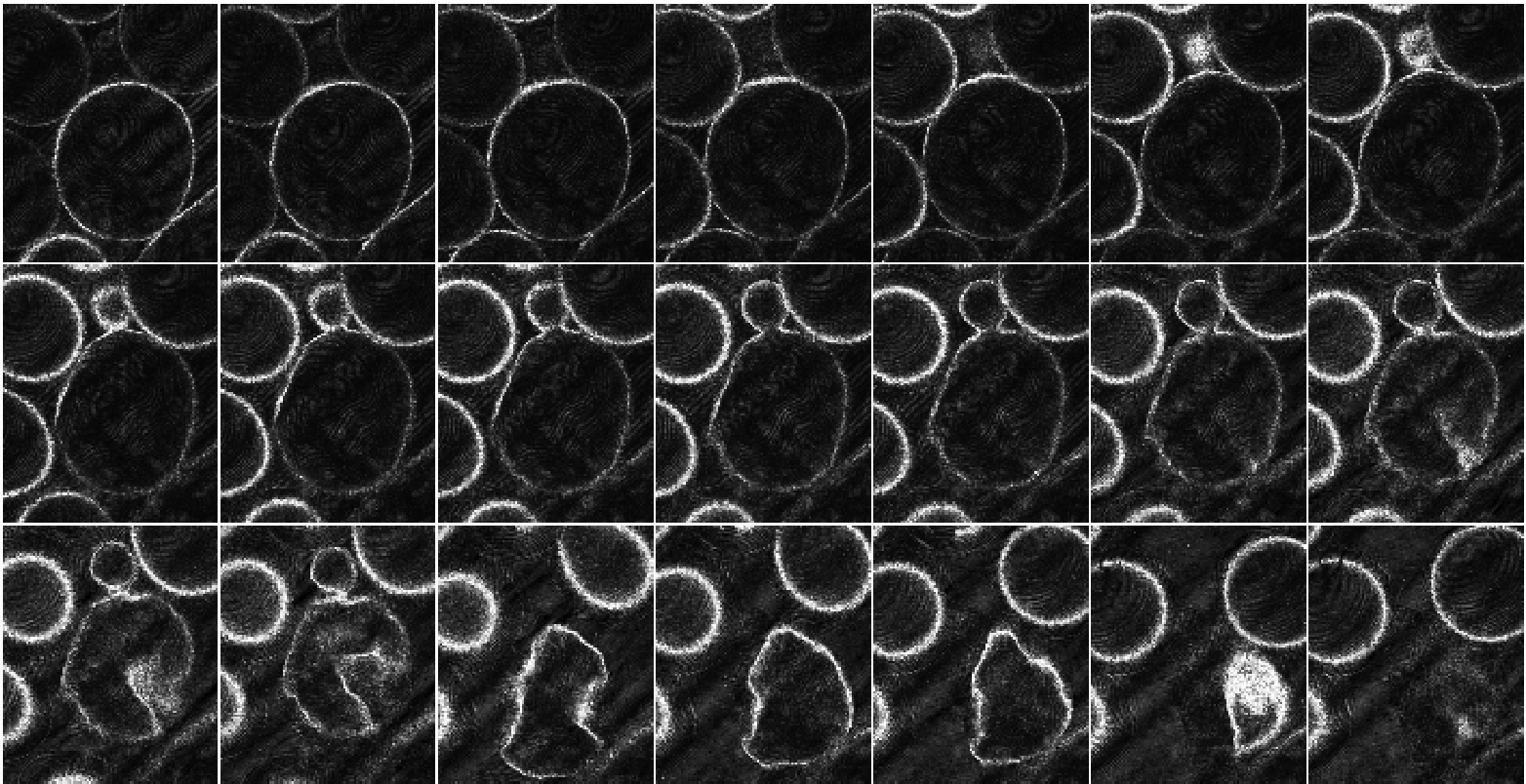}}
\caption{\label{Crumple}
Images of a heptane-in-methanol droplet deflating. Beginning in the top
left-hand corner, there is one
frame every 60 seconds with each 2$^{\circ}$C higher in temperature than
the previous. The bulk demixing temperature is just before the end of this 
sequence. The images are at a depth of 40 $\mu$m from the surface of the
holder and are roughly 300 $\mu$m on an edge.}
\end{figure}

To demonstrate that the fluid phase separation has given rigidity to the
particle-laden interfaces we have studied the behaviour as both the
interfacial
tension and the balance of volumes are varied.
We achieve this by slowly warming our particle-stabilized emulsions into
the mixed regime. Figure~\ref{Crumple} 
shows a series of frames including a central
droplet which becomes \textit{deflated}. This is a very clear change to
the behaviour shown in figure~\ref{Projection}(b \& c), 
for example. In figure~\ref{Crumple} the interfaces are
not stable and are rapidly changing in shape. An important point to note
is that most of the surrounding droplets are not crumpling and this
suggests that the integrity of the droplet surface for this individual
droplet has been lost. (The slow creaming of the droplets in these frames
indicates that none of them are stuck to the sample holder.)
The puncture in the particle skin allows the
minority fluid phase to leave the shell and mix with the continuous
phase.
A structure with a high bending modulus will tend to maintain smoothly
curved surfaces when it deflates. This phenomenon is well known from
studies of blood cells and some types of vesicles~\cite{Pamplona1996}. 
If the internal
pressure is reduced the structure of these 
can transform
into oblate shapes including a biconcave structure or prolate shapes
including dumbbells. This is not what we see on warming this sample.
Instead the shape we observe is puckered and creased.
This behaviour indicates that our particle-laden
interfaces are dominated by the energy cost of stretching the
interface. Similar crumpling behaviour is observed when a water-in-crude
oil droplet is deflated using a micropipette~\cite{Yeung1999}.

Other researchers have
studied the conditions
leading to crumpling of flat particle layers at liquid-liquid interfaces 
using a Langmuir
trough~\cite{Aveyard2000}.
It is found that a buckling transition
occurs when the inter-particle repulsion exactly balances the
interfacial
tension cost of expanding the region between particles. It seems likely
that a related transition is occurring here as we slowly reduce the 
interfacial tension and internal volume.

\begin{figure}
\centerline{\includegraphics[scale=0.6]{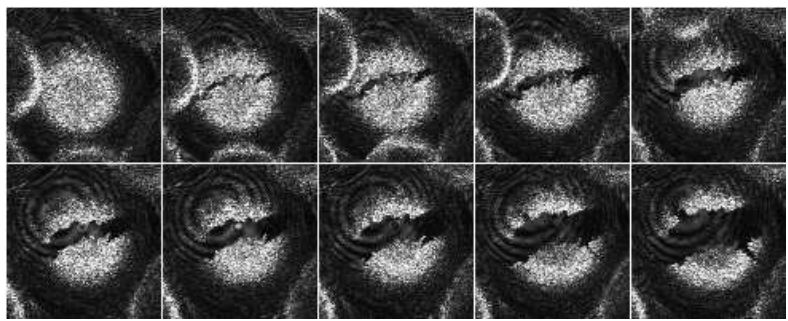}}
\caption{\label{Fracture}
Images of a heptane-in-methanol droplet cracking. Beginning in the top
left-hand corner, there is one
frame every 30 seconds with each 1$^{\circ}$C higher in temperature than
the previous. The bulk demixing temperature is in the middle of this 
sequence. The images are at a depth of 40 $\mu$m from the surface of the
holder and are roughly 120 $\mu$m on an edge.}
\end{figure}

We suggested above that puncturing of the droplet surface may have
led to the observed deflation. We have associated other observations
with droplets retaining their
integrity to higher temperatures. In figure~\ref{Fracture} it is shown that
droplets with a complete coverage of particles will eventually
shatter. This effect is often seen in the thin side walls of emulsion
droplets; however it is most visually dramatic to observe the fracture
in the end face of a droplet. The cracking reveals the solidity of these
particle layers. The layers are sufficiently rigid that they fracture
rather than distort.

\subsection{Coalescence and stability}

\begin{figure}
\centerline{\includegraphics[scale=0.9]{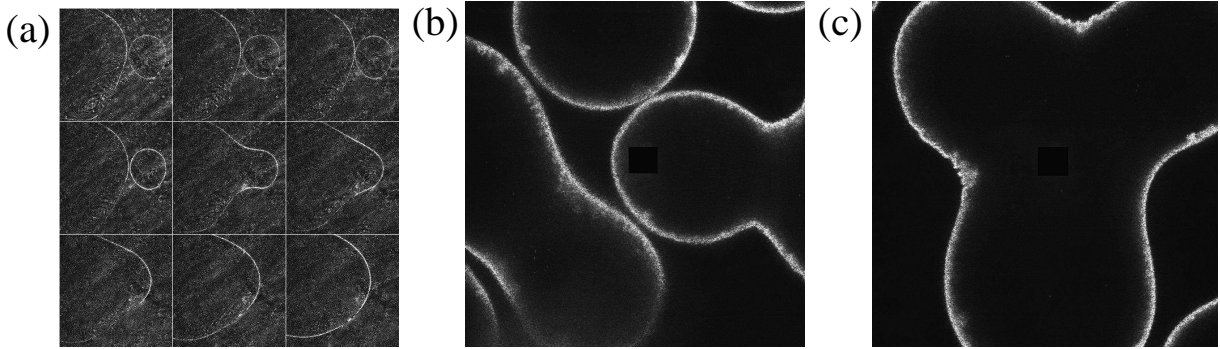}}
\caption{\label{Coalescence} 
(a) Coalescence in progress in a dodecane-in-ethanol emulsion. Beginning
in the top left-hand corner, there
is one frame (500 $\mu$m across) 
every 30 seconds with 1$^{\circ}$C between them.
(b \& c) Distorted droplets of heptane-in-methanol 
most likely due to coalescence during
cooling. The images are 200 $\mu$m across and are of different regions
of the same sample.}
\end{figure}

As the emulsions are heated toward miscibility coalescence becomes
increasingly common. Figure~\ref{Coalescence}(a) 
shows a smaller droplet coalescing with
its larger neighbour. The combined interface becomes smooth and round
within minutes. This contrasts strongly with behaviour at low
temperatures. Evidently in some cases the layer of colloids is becoming
sufficiently flexible and fluid for conventional coalescence behaviour
to occur. An interface of negative Gaussian curvature is stable for a
few tens of seconds (figure~\ref{Coalescence}(a) central panel). 

Such coalescence obviously can also occur during cooling.
Figure~\ref{Coalescence}(b \& c) 
shows two permanently distorted droplets. In both
cases there are regions of the interface with negative Gaussian
curvature.
The peanut and trefoil droplets
may well have fused
while the surfaces were fluid without having time to rearrange before
the bending and Young's moduli became too high.
It can be concluded that the large elastic moduli makes it impossible for
the droplet surface to change its shape. In addition the particle
interfaces are slightly wrinkled at the points of sharpest curvature in
figure~\ref{Coalescence}(c) 
and this suggests that the surface is solid and the particles
are unable to desorb.

We note that the existence of peanut/dumbbell shapes
(figure~\ref{Coalescence}(b \& c)) does not by itself demonstrate that the
particle surfaces are solid. Similar shapes are predicted for
vesicles with fluid surfaces within both the spontaneous-curvature and
bilayer-coupling models~\cite{Pamplona1996}~\cite{Seifert1991}. Instead
it is the crumpling and fracture behaviour which are the key to
understanding the nature of these emulsions.

\section{Summary}

We have presented confocal microscopy images of droplet geometries in
particle-stabilized emulsions. We have shown how the behaviour
dramatically changes as the fluids slowly become miscible. The droplets created
in our experiments have
surface layers of particles that are solid and have relatively
high elastic moduli. At low temperature the system is essentially
arrested and can make no attempt to rearrange itself on experimentally
relevant timescales.

Our results begin to reveal the conditions under which the coalescence of
droplets can occur. Figure~\ref{Projection}(b \& c), 
figure~\ref{Crumple} (upper images) and figure~\ref{Coalescence}(b \& c)
show
particle-stabilized droplets which have been in close contact for hours. It
is clear that the droplets are showing no tendency toward coalescing.
By contrast figure~\ref{Coalescence}(a) 
shows coalescence in action. It appears that
fluidity of the layers is essential before two droplets can coalesce.
This is an important observation concerning why particle-stabilized
monolayers have longevity.
The results presented here are qualitative. We intend to follow up our
observations with a study of the rheology of dense layers of silica
particles at liquid-liquid interfaces~\cite{Stancik2004}.

The creation route for these emulsions has involved a quench from high
temperature into the demixed region of the phase diagram. It is an
interesting extension of this research to explore the behaviour at higher
quench rates to lower temperatures. One approach would be to see how the
particles respond to different separation kinetics and the resulting
structures could be highly novel. These studies are underway.

\section{Acknowledgments}
We thank S.U.~Egelhaaf and D.~Roux for illuminating
discussions. Funding in Edinburgh was provided by the EPSRC (Grant
GR/S10377/01).


\vspace{0.5cm}

\begin{thebibliography}{100}
\bibitem{Aveyard2003}
R.~Aveyard, B.P.~Binks and J.H.~Clint 2003 \textit{Adv. Colloid Interface Sci.}
\textbf{100-102} 503

\bibitem{Rowlinson1982}
J.S.~Rowlinson and F.L.~Swinton  1982 \textit{Liquids and Liquid Mixtures}
3$^{rd}$ edition (Butterworths, London)

\bibitem{Stober1968}
W.~St\"{o}ber, A.~Fink and E.~Bohn 1968 \textit{J. Colloid Interface
Sci.} \textbf{26} 62

\bibitem{Horozov2003}
T.S.~Horozov, R.~Aveyard, J.H.~Clint and B.P.~Binks 2003 \textit{Langmuir}
\textbf{19} 2822

\bibitem{ImageJ}
W.S.~Rasband 1997-2005 \textit{ImageJ} (U.S. National Institute of Health, 
Bethesda,
Maryland, USA) http://rsb.info.nih.gov/ij/

\bibitem{Aveyard2000}
R.~Aveyard, J.H.~Clint, D.~Nees and N.~Quirke 2000 \textit{Langmuir} 
\textbf{16} 8820

\bibitem{Stancik2004}
E.J.~Stancik, A.L.~Hawkinson, J.~Vermant and G.G.~Fuller 2004 
\textit{J. Rheol.} \textbf{48} 159

\bibitem{Dickinson1999}
E.~Dickinson 1999 \textit{Colloids Surf. B} \textbf{15} 161

\bibitem{Evans1999}
D.F.~Evans and H.~Wennerstr\"{o}m 1999 \textit{The Colloidal Domain}
2$^{nd}$ edition (Wiley-VCH, New York)

\bibitem{Pamplona1996}
D.C.~Pamplona and C.R.~Calladine 1996 \textit{J. Biomech. Eng.} 
\textbf{118} 482

\bibitem{Yeung1999}
A.~Yeung, T.~Dabros, J.~Czarnecki and J.~Masliyah 1999 \textit{Proc. R.
Soc. Lond. A} \textbf{455} 3709

\bibitem{Seifert1991}
U.~Seifert, K.~Berndl and R.~Lipowsky 1991 \textit{Phys. Rev. A} 
\textbf{44} 1182
\end{thebibliography}
\end{document}